\documentclass[aps,preprint]{revtex4}%
\usepackage{amsfonts}
\usepackage{amsmath}
\usepackage{amssymb}
\usepackage{graphicx}%
\begin{document}
\preprint{CTP-SCU/2015004}
\title{Minimal Length Effects on Entanglement Entropy of Spherically Symmetric Black
Holes in Brick Wall Model}
\author{Peng Wang}
\email{pengw@scu.edu.cn}
\author{Haitang Yang}
\email{hyanga@scu.edu.cn}
\author{Shuxuan Ying}
\email{2013222020007@stu.scu.edu.cn}
\affiliation{Center for Theoretical Physics, College of Physical Science and Technology,
Sichuan University, Chengdu, 610064, PR China}

\begin{abstract}
We compute the black hole horizon entanglement entropy for a massless scalar
field in the brick wall model by incorporating the minimal length. Taking the
minimal length effects on the occupation number $n\left(  \omega,l\right)  $
and the Hawking temperature into consideration, we obtain the leading UV
divergent term and the subleading logarithmic term in the entropy. The leading
divergent term scales with the horizon area. The subleading logarithmic term
is the same as that in the usual brick wall model without the minimal length.\ 

\end{abstract}
\maketitle
\tableofcontents



\section{Introduction}

Bekenstein and Hawking showed that the entropy of a black hole is proportional
to the area of the
horizon\cite{IN-Bekenstein:1972tm,IN-Bekenstein:1973ur,IN-Hawking:1976de}.
Although all the evidences suggest that the Bekenstein-Hawking entropy is
truly the thermodynamic entropy, the statistical origin of the black-hole
entropy has not yet been fully understood. It seems that an unavoidable
candidate for the statistical origin is the entropy of the thermal atmosphere
of the black hole, which can also be thought of as the entanglement entropy
across the horizon\cite{IN-Bombelli:1986rw}.

To be generic, we will consider a spherically symmetric background metric of
the form
\begin{equation}
ds^{2}=f\left(  r\right)  dt^{2}-\frac{dr^{2}}{f\left(  r\right)  }%
-r^{2}\left(  d\theta^{2}+\sin^{2}\theta d\phi^{2}\right)  ,
\label{eq:SchwarzschildMetric}%
\end{equation}
where $f\left(  r\right)  $ has a simple zero at $r=r_{h}$ with $f^{\prime
}\left(  r_{h}\right)  $ being finite and nonzero. The vanishing of $f\left(
r\right)  $ at point $r=r_{h}$ indicates the presence of an event horizon.
Thus, the atmosphere entropy of the black hole with the metric $\left(
\ref{eq:SchwarzschildMetric}\right)  $ can be expressed in the form%
\begin{equation}
S=\sum_{l=0}^{\infty}\left(  2l+1\right)  \int dn\left(  \omega,l\right)
s\left(  \frac{\omega}{T}\right)  , \label{eq:EntropySum}%
\end{equation}
where $\omega$ is the Killing energy associated with $t$, $l$ is the angular
momentum, $T$ is the Hawking temperature of the black hole, $n\left(
\omega,l\right)  $ is the number of one-particle states not exceeding $\omega$
with fixed value of angular momentum $l$, and $s\left(  \frac{\omega}%
{T}\right)  $ is the thermal entropy per mode. However, the entropy diverges
when we attempt to sum eqn. $\left(  \ref{eq:EntropySum}\right)  $ over all
the modes. There are two kinds of divergences. The first one is due to
infinite volume of the system, which has to do with the contribution from the
vacuum surrounding the system at large distances and is of little relevance
here. The second one arises from the infinite volume of the deep throat region
near the horizon. In order to regulate the two divergences, t' Hooft
\cite{IN-'tHooft:1984re}\ proposed the brick wall model for a scalar field
$\phi$, where two brick wall cutoffs are introduced at some small distance
$r_{\varepsilon}$ from the horizon and at a large distance $L\gg r_{h}$,
\begin{equation}
\phi=0\text{ \ at }r=r_{h}+r_{\varepsilon}\text{ and }r=L\text{.}
\label{eq:DirichletBoundary}%
\end{equation}
The minimally coupled scalar field satisfies the Klein-Gordon equation%
\begin{equation}
\left(  \frac{\nabla^{2}}{\hbar^{2}}-m^{2}\right)  \phi=0.
\label{eq:KleinGordon}%
\end{equation}
Since WKB approximation is reliable as long as the black hole's mass $M\gg
1$(in Planck units), t' Hooft took the ansatz for $\phi$
\begin{equation}
\phi=\exp\left[  -\frac{i}{\hbar}\omega t+\frac{i}{\hbar}\int p_{r}dr\right]
Y_{lm}\left(  \theta,\phi\right)  ,
\end{equation}
and solved for $p_{r}$ to the WKB leading order. Define the radial wave number
$k\left(  r,l,\omega\right)  $ by%
\begin{equation}
k\left(  r,l,\omega\right)  =\left\vert p_{r}\right\vert , \label{eq:K}%
\end{equation}
as long as $p_{r}^{2}\geq0$, and $k\left(  r,l,\omega\right)  =0$ otherwise.
With the two Dirichlet boundaries $\left(  \ref{eq:DirichletBoundary}\right)
$, $n\left(  \omega,l\right)  $ can be expressed as%
\begin{equation}
n\left(  \omega,l\right)  =\frac{1}{\pi\hbar}\int_{r_{h}+r_{\varepsilon}}%
^{L}k\left(  r,l,\omega\right)  dr. \label{eq:N}%
\end{equation}

It would appear that the entanglement entropy is sensitive to the
ultraviolet(UV) behavior of quantum fields, where quantum gravity effects
become important. Even though there is still no complete and consistent
quantum theory of gravity, one could still rely on effective models to study
the UV behavior of the entanglement entropy. For example, this issue was
studied in \cite{IN-Chang:2003sa,IN-Jacobson:2007jx} in the context of the
possible modifications of the standard dispersion relation. On the other hand,
various theories of quantum gravity, such as string theory, loop quantum
gravity and quantum geometry, predict the existence of a minimal
length\cite{IN-Townsend:1977xw,IN-Amati:1988tn,IN-Konishi:1989wk}. The
generalized uncertainty principle(GUP)\cite{IN-Kempf:1994su} is a simple way
to realize this minimal length. In
\cite{IN-Li:2002xb,IN-Kim:2006rx,IN-Kim:2007bx,IN-Kim:2007nh}, the authors
considered the generalized uncertainty relation%
\begin{equation}
\Delta x\Delta p\geq\hbar+\frac{\lambda}{\hbar}\left(  \Delta p\right)  ^{2},
\label{eq:LiForm}%
\end{equation}
which gives a minimal length, $2\sqrt{\lambda}$. As a consequence, the number
of quantum states should be changed to
\begin{equation}
\frac{d^{3}xd^{3}p}{\left(  2\pi\hbar\right)  ^{3}\left(  1+\lambda
p^{2}\right)  ^{3}},
\end{equation}
where $p^{2}=p_{i}p^{i}$. Therefore, they had for massless particles%
\begin{equation}
n\left(  \omega,l\right)  =\frac{1}{\pi\hbar}\int_{r_{h}+r_{\varepsilon}}%
^{L}\frac{k\left(  r,l,\omega\right)  }{\left(  1+\lambda\omega^{2}/f\right)
^{3}}dr, \label{eq:NLamda}%
\end{equation}
where $k\left(  r,l,\omega\right)  $ is given by eqn. $\left(  \ref{eq:K}%
\right)  $. The all order generalized uncertainty relation was also considered
in \cite{IN-Kim:2007if}. It is found there that the artificial cutoff
parameter in the brick wall model located just outside the horizon can be
avoided if GUP is considered. Alternatively, the modified Klein-Gordon
equation for $\phi$ was considered in the framework of Horava-Lifshitz gravity
and GUP in \cite{IN-Eune:2010kx}. The WKB leading term of $p_{r}$ for the
modified Klein-Gordon equation was obtained and $n\left(  \omega,l\right)  $
was given by eqn. $\left(  \ref{eq:N}\right)  $. The entanglement entropy was
then calculated and the result is consistent with previous studies.

Furthermore, GUP should modify the Hawking temperature $T$ in eqn. $\left(
\ref{eq:EntropySum}\right)  $ as well as $n\left(  \omega,l\right)  $. Indeed,
the GUP deformed Hamilton-Jacobi equation in curved spacetime have been
introduced and the corrected Hawking temperatures have been derived in
\cite{IN-Chen:2013pra,IN-Chen:2013tha,IN-Chen:2013ssa,IN-Chen:2014xsa,IN-Mu:2015qta}%
. The GUP corrections to the Hawking temperature were found to depend not only
on the black hole's mass but the mass, the energy, and the angular momentum of
emitted particles as well. In \cite{IN-Mu:2015qta}, we derived the deformed
Klein-Gordon and Dirac equation incorporating the GUP form proposed in
\cite{DHM-Hossenfelder:2003jz,DHM-Hassan:2002qk}. The GUP modified Hawking
temperatures for scalars and fermions were then obtained and the black hole's
evaporation was also discussed in \cite{IN-Mu:2015qta}. In this paper, we will
investigate the entropy of the thermal atmosphere of the black hole in the
context of \cite{IN-Mu:2015qta}. Taking GUP corrections to both $n\left(
\omega,l\right)  $ and Hawking temperature $T$ into consideration, we will
calculate the entanglement entropy of a massless scalar field via eqn.
$\left(  \ref{eq:EntropySum}\right)  $ in the brick wall model.

The organization of this paper is as follows. In section \ref{Sec:DHM}, we
review some results of \cite{IN-Mu:2015qta}, which are necessary for
calculating the entanglement entropy. In section \ref{Sec:EEBH}, the
entanglement entropy of a massless scalar field near the horizon is calculated
in the brick wall model. Section \ref{Sec:Con} is devoted to our discussion
and conclusion. In this paper, we take Geometrized units $c=G=1$, where the
Planck constant $\hbar$ is square of the Planck Mass $m_{p}$. For simplicity,
we assume that the emitted particles are massless and neutral.

\section{Deformed Hamilton-Jacobi Method}

\label{Sec:DHM}

In \cite{IN-Mu:2015qta} and this paper, we consider an effective model of the
GUP in one dimensional quantum mechanics given
by\cite{DHM-Hossenfelder:2003jz,DHM-Hassan:2002qk}
\begin{equation}
L_{f}k(p)=\tanh\left(  \frac{p}{M_{f}}\right)  , \label{tanh1}%
\end{equation}%
\begin{equation}
L_{f}\omega(E)=\tanh\left(  \frac{E}{M_{f}}\right)  , \label{tanh2}%
\end{equation}
where the generators of the translations in space and time are the wave vector
$k$ and the frequency $\omega$, $L_{f}$ is the minimal length, and $L_{f}%
M_{f}=\hbar$. From eqn. $\left(  \ref{tanh1}\right)  $, it is noted that
although one can increase $p$ arbitrarily, $k$ has an upper bound which is
$\frac{1}{L_{f}}$. The upper bound on $k$ implies that particles could not
possess arbitrarily small Compton wavelength $\lambda=2\pi/k$ and that there
exists a minimal length $\sim L_{f}$. The quantization in position
representation $\hat{x}=x$ leads to
\begin{equation}
k=-i\partial_{x},\text{ }\omega=i\partial_{t}.
\end{equation}
In the $\left(  3+1\right)  $ dimensional flat spacetime, the relations
between $\left(  p_{i},E\right)  $ and $\left(  k_{i},\omega\right)  $ can
simply be generalized to%
\begin{align}
L_{f}k_{i}(p)  &  =\tanh\left(  \frac{p_{i}}{M_{f}}\right)
,\label{eq:wavevector}\\
L_{f}\omega(E)  &  =\tanh\left(  \frac{E}{M_{f}}\right)  ,
\label{eq:frequency}%
\end{align}
where one has for $\vec{k}$ in the spherical coordinates%
\begin{equation}
\vec{k}=-i\left(  \hat{r}\frac{\partial}{\partial r}+\frac{\hat{\theta}}%
{r}\frac{\partial}{\partial\theta}+\frac{\hat{\phi}}{r\sin\theta}%
\frac{\partial}{\partial\phi}\right)  .
\end{equation}
Expanding eqn. $\left(  \ref{eq:wavevector}\right)  $ and eqn. $\left(
\ref{eq:frequency}\right)  $ for small arguments to the third order gives the
energy and momentum operator in position representation%
\begin{gather}
E=i\hbar\partial_{t}\left(  1-\frac{\hbar^{2}}{M_{f}^{2}}\partial_{t}%
^{2}\right)  ,\\
\vec{p}=\frac{\hbar}{i}\left[  \hat{r}\left(  \partial_{r}-\frac{\hbar
^{2}\partial_{r}^{3}}{M_{f}^{2}}\right)  +\hat{\theta}\left(  \frac
{\partial_{\theta}}{r}-\frac{\hbar^{2}\partial_{\theta}^{3}}{r^{3}M_{f}^{2}%
}\right)  +\hat{\phi}\left(  \frac{\partial_{\phi}}{r\sin\theta}-\frac
{\hbar^{2}\partial_{\phi}^{3}}{r^{3}\sin^{3}\theta M_{f}^{2}}\right)  \right]
,
\end{gather}
where we also omit the factor $\frac{1}{3}$. Substituting the above energy and
momentum operators into the energy-momentum relation, the deformed
Klein-Gordon equation satisfied by the massless scalar field is
\begin{equation}
E^{2}\phi=p^{2}\phi, \label{eq:KGFD}%
\end{equation}
where $p^{2}=\vec{p}\cdot\vec{p}.$ Making the ansatz for $\phi$
\begin{equation}
\phi=\exp\left(  \frac{iI}{\hbar}\right)  ,
\end{equation}
and substituting it into eqn. $\left(  \ref{eq:KGFD}\right)  $, one expands
eqn. $\left(  \ref{eq:KGFD}\right)  $ in powers of $\hbar$ and then finds that
the lowest order gives the deformed scalar Hamilton-Jacobi equation in the
flat spacetime
\begin{gather}
\left(  \partial_{t}I\right)  ^{2}\left(  1+\frac{2\left(  \partial
_{t}I\right)  ^{2}}{M_{f}^{2}}\right)  -\left(  \partial_{r}I\right)
^{2}\left(  1+\frac{2\left(  \partial_{r}I\right)  ^{2}}{M_{f}^{2}}\right)
-\frac{\left(  \partial_{\theta}I\right)  ^{2}}{r^{2}}\left(  1+\frac{2\left(
\partial_{\theta}I\right)  ^{2}}{r^{2}M_{f}^{2}}\right) \nonumber\\
-\frac{\left(  \partial_{\phi}I\right)  ^{2}}{r^{2}\sin^{2}\theta}\left(
1+\frac{2\left(  \partial_{\phi}I\right)  ^{2}}{r^{2}\sin^{2}\theta M_{f}^{2}%
}\right)  =0, \label{eq:HJFD}%
\end{gather}
which is truncated at $\mathcal{O}\left(  \frac{1}{M_{f}^{2}}\right)  $. The
Hamilton-Jacobi equation in the metric $\left(  \ref{eq:SchwarzschildMetric}%
\right)  $ can be obtained from that in flat spacetime by making replacements
$\partial_{r}I\rightarrow\sqrt{f\left(  r\right)  }\partial_{r}I$ and
$\partial_{t}I\rightarrow\frac{\partial_{t}I}{\sqrt{f\left(  r\right)  }}$.
Therefore, the deformed Hamilton-Jacobi equation in flat spacetime, eqn.
$\left(  \ref{eq:HJFD}\right)  $, leads to the deformed Hamilton-Jacobi
equation in the metric $\left(  \ref{eq:SchwarzschildMetric}\right)  $, which
is to $\mathcal{O}\left(  \frac{1}{M_{f}^{2}}\right)  $
\begin{gather}
\frac{\left(  \partial_{t}I\right)  ^{2}}{f\left(  r\right)  }\left(
1+\frac{2\left(  \partial_{t}I\right)  ^{2}}{f\left(  r\right)  M_{f}^{2}%
}\right)  -f\left(  r\right)  \left(  \partial_{r}I\right)  ^{2}\left(
1+\frac{2f\left(  r\right)  \left(  \partial_{r}I\right)  ^{2}}{M_{f}^{2}%
}\right)  -\frac{\left(  \partial_{\theta}I\right)  ^{2}}{r^{2}}\left(
1+\frac{2\left(  \partial_{\theta}I\right)  ^{2}}{r^{2}M_{f}^{2}}\right)
\nonumber\\
-\frac{\left(  \partial_{\phi}I\right)  ^{2}}{r^{2}\sin^{2}\theta}\left(
1+\frac{2\left(  \partial_{\phi}I\right)  ^{2}}{r^{2}\sin^{2}\theta M_{f}^{2}%
}\right)  =0. \label{eq:HJCG}%
\end{gather}

Taking into account the Killing vectors of the background spacetime, we can
employ the following ansatz for the action
\begin{equation}
I=-\omega t+W\left(  r,\theta\right)  +p_{\phi}\phi, \label{eq:Ianzatz}%
\end{equation}
where $\omega$ and $p_{\phi}$ are constants and they are the energy and the
$z$-component of angular momentum of emitted particles, respectively.
Inserting eqn. $(\ref{eq:Ianzatz})$ into eqn. $(\ref{eq:HJCG}),$ we find that
the deformed Hamilton-Jacobi equation becomes
\begin{equation}
p_{r}^{2}\left(  1+\frac{2f\left(  r\right)  p_{r}^{2}}{M_{f}^{2}}\right)
=\frac{1}{f^{2}\left(  r\right)  }\left[  \omega^{2}\left(  1+\frac
{2\omega^{2}}{f\left(  r\right)  M_{f}^{2}}\right)  -f\left(  r\right)
\frac{L^{2}}{r^{2}}\right]  , \label{eq:pr}%
\end{equation}
where we neglect terms higher than $\mathcal{O}\left(  \frac{1}{M_{f}^{2}%
}\right)  $. In the above equation, we define $p_{r}=\partial_{r}W$,
$p_{\theta}=\partial_{\theta}W$, $L^{2}=p_{\theta}^{2}+\frac{p_{\phi}^{2}%
}{\sin^{2}\theta}$, and $L^{2}=\left(  l+1\right)  l\hbar^{2}$ which is the
magnitude of the angular momentum of the particle. Solving eqn. $(\ref{eq:pr}%
)$ for $p_{r}$ to $\mathcal{O}\left(  \frac{1}{M_{f}^{2}}\right)  $gives%
\begin{equation}
p_{r}^{\pm}=\pm\left(  \frac{\Omega^{\frac{1}{2}}}{f\left(  r\right)  }%
-\frac{\Omega^{\frac{3}{2}}}{f^{2}\left(  r\right)  M_{f}^{2}}\right)  ,
\label{eq:PrSolution}%
\end{equation}
where +/$-$ represent the outgoing/ingoing solutions and we define
\begin{equation}
\Omega=\omega^{2}\left(  1+\frac{2\omega^{2}}{f\left(  r\right)  M_{f}^{2}%
}\right)  -f\left(  r\right)  \frac{\left(  l+1\right)  l\hbar^{2}}{r^{2}}.
\end{equation}
Using the residue theory for semi circles, we obtain for the imaginary part of
$W_{\pm}$ to $\mathcal{O}\left(  \frac{1}{M_{f}^{2}}\right)  $
\begin{equation}
\operatorname{Im}W_{\pm}=\pm\frac{\pi\omega}{f^{\prime}\left(  r_{h}\right)
}\left[  1+\frac{2\left(  l+1\right)  l\hbar^{2}}{M_{f}^{2}r_{h}^{2}}\right]
.
\end{equation}
Taking both the spatial contribution and the temporal contribution into
account\cite{DHM-Akhmedova:2008au,DHM-Akhmedova:2008dz,DHM-Akhmedov:2008ru,DHM-Chowdhury:2006sk,DHM-Akhmedov:2006un}%
, one finds that the tunnelling rate of the particle crossing the horizon is
\begin{align}
\Gamma &  \propto exp\left[  -\frac{1}{\hbar}\left(  \operatorname{Im}\left(
\omega\Delta t\right)  +\operatorname{Im}\oint p_{r}dr\right)  \right]
\nonumber\\
&  =exp\left\{  -\frac{4\pi\omega}{f^{\prime}\left(  r_{h}\right)  }\left[
1+\frac{2\left(  l+1\right)  l\hbar^{2}}{M_{f}^{2}r_{h}^{2}}\right]  \right\}
. \label{eq:Gamma}%
\end{align}
This is the expression of Boltzmann factor with an effective temperature to
$\mathcal{O}\left(  \frac{1}{M_{f}^{2}}\right)  $
\begin{equation}
T=\frac{T_{0}}{1+\Delta}, \label{HKT1}%
\end{equation}
where $T_{0}=\frac{\hbar f^{\prime}\left(  r_{h}\right)  }{4\pi}$ is the
original Hawking temperature and we define%
\begin{equation}
\Delta=\frac{\left(  l+1\right)  l\hbar^{2}}{M_{f}^{2}r_{h}^{2}}.
\label{eq:Delta}%
\end{equation}
.

\section{Entanglement Entropy of Black Hole}

\label{Sec:EEBH}

For scalar particles emitted in a wave mode labelled by energy $\omega$ and
$l,$ we find from eqn. $\left(  \ref{eq:Gamma}\right)  $ that
\cite{EEBH-Hartle:1976tp}%
\begin{align*}
&  \left(  \text{Probability for a black hole to emit a particle in this
mode}\right) \\
&  =\exp\left(  -\frac{\omega}{T}\right)  \times(\text{Probability for a black
hole to absorb a particle in the same mode}),
\end{align*}
where $T$ is given by eqn. $\left(  \ref{HKT1}\right)  $. Neglecting
back-reaction, detailed balance condition requires that the ratio of the
probability of having $N$ particles in a particular mode with $\omega$ and $l$
to the probability of having $N-1$ particles in the same mode is $\exp\left(
-\frac{\omega}{T}\right)  .$ One then follows the standard textbook procedure
to get the average number $n_{\omega,l}$ in the mode%
\begin{equation}
n_{\omega,l}=n\left(  \frac{\omega}{T}\right)  ,
\end{equation}
where we define%
\begin{equation}
n\left(  x\right)  =\frac{1}{\exp x-1}.
\end{equation}
The von Neumann entropy for the mode is%
\begin{equation}
s_{\omega,l}=\left(  n_{\omega,L}+1\right)  \ln\left(  1+n_{\omega,L}\right)
-n_{\omega,L}\ln n_{\omega,L}, \label{eq:EntropyForMode}%
\end{equation}
where $s_{\omega,l}$ is $s\left(  \frac{\omega}{T}\right)  $ in eqn. $\left(
\ref{eq:EntropySum}\right)  $. The $s\left(  x\right)  $ is given by%
\begin{equation}
s\left(  x\right)  =\frac{\exp x}{\exp x-1}\ln\left(  \frac{\exp x}{\exp
x-1}\right)  +\frac{\ln\left(  \exp x-1\right)  }{\exp x-1}.
\end{equation}

Define the radial wave number $k\left(  r,l,\omega\right)  $ by%
\begin{equation}
k\left(  r,l,\omega\right)  =\left\vert p_{r}^{\pm}\right\vert ,
\end{equation}
as long as $p_{r}^{2}\geq0$, and $k\left(  r,l,\omega\right)  =0$ otherwise.
The $p_{r}^{\pm}$ is given by eqn. $\left(  \ref{eq:PrSolution}\right)  $.
Taking two Dirichlet conditions at $r=r_{h}+r_{\varepsilon}$ and $r=L$ in the
brick wall model into account, we find that the number of one-particle states
not exceeding $\omega$ with fixed value of angular momentum $l$ is given by%
\begin{equation}
n\left(  \omega,l\right)  =\frac{1}{\pi\hbar}\int_{r_{h}+r_{\varepsilon}}%
^{L}k\left(  r,l,\omega\right)  dr.
\end{equation}
Thus, we get for the total entropy of radiation%
\begin{align}
S  &  =\sum\limits_{\omega,l,m}s_{\omega,l}=\int\left(  2l+1\right)  dl\int
d\omega\frac{dn\left(  \omega,l\right)  }{d\omega}s_{\omega,l}\nonumber\\
&  =-\frac{1}{\pi\hbar^{3}}\int d\left[  \left(  l+1\right)  l\hbar
^{2}\right]  \int d\omega\frac{\partial s_{\omega,l}}{\partial\omega}%
\int_{r_{h}+r_{\varepsilon}}^{L}drk\left(  r,l,\omega\right)  .
\end{align}
Defining $u=\frac{\omega}{T_{0}}$ and expanding $s_{\omega,l}$ to
$\mathcal{O}\left(  \frac{1}{M_{f}^{2}}\right)  $%
\begin{equation}
s_{\omega,l}\approx s\left(  u\right)  +s^{\prime}\left(  u\right)  u\Delta,
\end{equation}
we find that the entropy to $\mathcal{O}\left(  \frac{1}{M_{f}^{2}}\right)  $
is
\[
S\approx-\frac{M_{f}^{2}r_{h}^{2}}{\pi\hbar^{3}}\int du\int_{r_{h}%
+r_{\varepsilon}}^{L}dr\int d\Delta\left[  s^{\prime}\left(  u\right)
+\left(  s^{\prime}\left(  u\right)  u\right)  ^{\prime}\Delta\right]  \left(
\frac{\Omega^{\frac{1}{2}}}{f\left(  r\right)  }-\frac{\Omega^{\frac{3}{2}}%
}{f^{2}\left(  r\right)  M_{f}^{2}}\right)  ,
\]
where we use eqn. $\left(  \ref{eq:Delta}\right)  $ for $\Delta$ and eqn.
$\left(  \ref{eq:PrSolution}\right)  $ for $k\left(  r,l,\omega\right)  $.
Performing $\Delta$ integral which runs over the region where $\Omega>0$ gives%
\begin{equation}
S\approx\frac{2r_{h}^{2}T_{0}^{3}}{\pi\hbar^{3}}\int s\left(  u\right)
du\int_{r_{h}+r_{\varepsilon}}^{L}dr\frac{r^{2}}{r_{h}^{2}}\frac{u^{2}}%
{f^{2}\left(  r\right)  }\left[  1+\frac{4T_{0}^{2}u^{2}}{f\left(  r\right)
M_{f}^{2}}-\frac{6T_{0}^{2}u^{2}}{5f\left(  r\right)  M_{f}^{2}}\frac{r^{2}%
}{r_{h}^{2}}\right]  ,
\end{equation}
where the second term in the bracket comes from the GUP corrections to
$n\left(  \omega,l\right)  $ and the third term from the GUP corrections to
Hawking temperature $T$. Focusing on the divergent parts near horizon, we find%
\begin{align}
\int_{r_{h}+r_{\varepsilon}}^{L}dr\frac{r^{2}}{r_{h}^{2}}\frac{1}{f^{2}\left(
r\right)  }  &  \sim\frac{1}{4\kappa^{2}r_{\varepsilon}}-\left(  \frac
{1}{\kappa r_{h}}-\eta\right)  \frac{\ln\kappa r_{\varepsilon}}{2\kappa
},\nonumber\\
\int_{r_{h}+r_{\varepsilon}}^{L}dr\frac{r^{2}}{r_{h}^{2}}\frac{1}{f^{3}\left(
r\right)  }  &  \sim\frac{1}{4\kappa^{3}r_{\varepsilon}^{2}}+\frac{\frac
{2}{\kappa r_{h}}-3\eta}{8\kappa^{2}r_{\varepsilon}}-\left(  \frac{1}%
{\kappa^{2}r_{h}^{2}}+3(2\eta^{2}-\theta)-\frac{6\eta}{\kappa r_{h}}\right)
\frac{\ln\kappa r_{\varepsilon}}{8\kappa},\nonumber\\
\int_{r_{h}+r_{\varepsilon}}^{L}dr\frac{r^{4}}{r_{h}^{4}}\frac{1}{f^{3}\left(
r\right)  }  &  \sim\frac{1}{4\kappa^{3}r_{\varepsilon}^{2}}+\frac{\frac
{4}{\kappa r_{h}}-3\eta}{8\kappa^{2}r_{\varepsilon}}-\left(  \frac{6}%
{\kappa^{2}r_{h}^{2}}+3(2\eta^{2}-\theta)-\frac{12\eta}{\kappa r_{h}}\right)
\frac{\ln\kappa r_{\varepsilon}}{8\kappa}, \label{eq:FIntegrals}%
\end{align}
where we expand $f\left(  r\right)  $ and $\frac{r^{2}}{r_{h}^{2}}$ at
$r=r_{h}$
\begin{align}
f\left(  r\right)   &  \sim2\kappa\left(  r-r_{h}\right)  \left[  1+\eta
\kappa\left(  r-r_{h}\right)  +\theta\kappa^{2}\left(  r-r_{h}\right)
^{2}\right]  ,\nonumber\\
\frac{r^{2}}{r_{h}^{2}}  &  \sim1+\frac{2\kappa\left(  r-r_{h}\right)
}{\kappa r_{h}}+\frac{\kappa^{2}\left(  r-r_{h}\right)  ^{2}}{\kappa^{2}%
r_{h}^{2}}. \label{eq:Expansion}%
\end{align}
In eqn. $\left(  \ref{eq:FIntegrals}\right)  ,\,$\ we neglect finite terms as
$\kappa r_{\varepsilon}\rightarrow0$ and terms involving $L$. Note that we
define $\kappa=\frac{f^{\prime}\left(  r_{h}\right)  }{2}$ which is the
surface gravity for the black hole and hence $T_{0}=\frac{\hbar\kappa}{2\pi}$.
Thus, the divergent part of entropy near the horizon is%
\begin{align}
S  &  \sim\frac{2r_{h}^{2}T_{0}^{3}}{\pi\hbar^{3}\kappa}\int u^{2}s\left(
u\right)  du\left\{  \frac{1}{4\kappa r_{\varepsilon}}-\left(  \frac{1}{\kappa
r_{h}}-\eta\right)  \frac{\ln\kappa r_{\varepsilon}}{2}+\right. \nonumber\\
&  \left.  \frac{T_{0}^{2}u^{2}}{M_{f}^{2}}\left[  \frac{14}{20}\frac
{1}{\kappa^{2}r_{\varepsilon}^{2}}+\frac{\frac{8}{\kappa r_{h}}-21\eta
}{20\kappa r_{\varepsilon}}+\frac{2}{5}\left(  \frac{28}{\kappa^{2}r_{h}^{2}%
}+39(2\eta^{2}-\theta)-\frac{96\eta}{\kappa r_{h}}\right)  \frac{\ln\kappa
r_{\varepsilon}}{8}\right]  \right\}  . \label{eq:EntropyDivergentR}%
\end{align}
Defining the proper distance distance $\varepsilon$ between the brick wall and
the horizon%
\[
\varepsilon=\int_{r_{h}}^{r_{h}+r_{\varepsilon}}\frac{dr}{\sqrt{f}}\sim
\frac{\sqrt{r_{\varepsilon}}}{\sqrt{2\kappa}}\left(  2-\frac{\eta\kappa
r_{\varepsilon}}{3}\right)  ,
\]
one could express eqn. $\left(  \ref{eq:EntropyDivergentR}\right)  $ in terms
of $\varepsilon$%
\begin{align}
S  &  \sim\frac{r_{h}^{2}}{8\pi^{4}}\int u^{2}s\left(  u\right)  du\left\{
\frac{1}{\varepsilon^{2}}-2\kappa^{2}\left(  \frac{1}{\kappa r_{h}}%
-\eta\right)  \ln\kappa\varepsilon\right. \nonumber\\
&  \left.  +\frac{L_{f}^{2}u^{2}\kappa^{2}}{2\pi^{2}\varepsilon^{2}}\left[
\frac{14}{5\kappa^{2}\varepsilon^{2}}+\frac{2\left(  5\eta-\frac{39}{\kappa
r_{h}}\right)  }{45}+\frac{2\kappa^{2}\varepsilon^{2}}{5}\left(  \frac
{28}{\kappa^{2}r_{h}^{2}}+39(2\eta^{2}-\theta)-\frac{96\eta}{\kappa r_{h}%
}\right)  \frac{\ln\kappa\varepsilon}{4}\right]  \right\}  ,
\end{align}
where we use $L_{f}=\frac{\hbar}{M_{f}}$. A natural choice for $\varepsilon$
is the minimal length $L_{f}$. If we take $\varepsilon=L_{f}$, we have for the
entanglement entropy near the horizon
\begin{align}
S  &  \sim\frac{r_{h}^{2}}{8\pi^{4}L_{f}^{2}}\int_{0}^{\infty}\left(
1+\frac{14u^{2}}{10\pi^{2}}\right)  u^{2}s\left(  u\right)  du\nonumber\\
&  \left.  -\frac{r_{h}\kappa}{4\pi^{4}}\left(  1-\eta\kappa r_{h}\right)
\int u^{2}s\left(  u\right)  du\ln\kappa L_{f}+\text{Finite terms as
}\varepsilon L_{f}\rightarrow0\right. \nonumber\\
&  \sim\frac{17A}{1800\pi L_{f}^{2}}-\frac{r_{h}\kappa\left(  1-\eta\kappa
r_{h}\right)  }{45}\ln\kappa L_{f}+\text{Finite terms as }\varepsilon
L_{f}\rightarrow0, \label{eq:EntropyD}%
\end{align}
where $A=4\pi r_{h}^{2}$ is the horizon area.

\section{ Discussion and Conclusion}

\label{Sec:Con}

In \cite{IN-'tHooft:1984re,CON-Solodukhin:2011gn}, the entanglement entropy of
a massless scalar field near the horizon of the metric $\left(
\ref{eq:SchwarzschildMetric}\right)  $ was calculated in the brick wall model
without the GUP. It has been shown the entropy in this case is%
\begin{equation}
S\sim\frac{A}{360\pi\varepsilon^{2}}-\frac{\kappa r_{h}}{45}\left(
1-\kappa\eta r_{h}\right)  \ln\kappa\varepsilon, \label{eq:EntropyNo}%
\end{equation}
where $\kappa$ and $\eta$ are defined in eqn. $\left(  \ref{eq:Expansion}%
\right)  $ and $\varepsilon$ is the proper distance between the wall and the
horizon. In our paper, we take $\varepsilon=L_{f}$ which is the minimal
length. The numerical factor of the leading term of the entropy in eqn.
$\left(  \ref{eq:EntropyD}\right)  $ has been changed from $\frac{1}{360\pi}$
to $\frac{17}{1800\pi}$ due to the minimal length effects. The leading terms
of eqn. $\left(  \ref{eq:EntropyD}\right)  $ and eqn. $\left(
\ref{eq:EntropyNo}\right)  $ are both proportional to the horizon area. The
area law in the brick wall model was first obtained in
\cite{IN-'tHooft:1984re} and later was also studied in
\cite{CON-Mann:1990fk,CON-Susskind:1994sm,CON-Barbon:1994ej,CON-Mukohyama:1998rf}%
. In \cite{IN-Li:2002xb} where the GUP form $\left(  \ref{eq:LiForm}\right)  $
was used, the leading term of the entanglement entropy for the spherical
metric $\left(  \ref{eq:SchwarzschildMetric}\right)  $ was calculated up to
$\mathcal{O}\left(  \lambda\right)  $. The result was
\begin{equation}
S\sim\frac{3}{L_{f}^{2}\pi}\frac{A}{4},
\end{equation}
where $L_{f}=2\sqrt{\lambda}$ is the minimal length. In \cite{IN-Kim:2007if}
where the all order GUP form
\begin{equation}
\Delta x\Delta p\geq\frac{\hbar}{2}\exp\left[  \frac{\lambda}{\hbar^{2}%
}\left(  \Delta p\right)  ^{2}\right]  ,
\end{equation}
was considered, the leading term of the all order entanglement entropy for the
spherically symmetric metric $\left(  \ref{eq:SchwarzschildMetric}\right)  $
was given by%
\begin{equation}
S\sim\frac{e^{3}\zeta\left(  3\right)  }{4\pi L_{f}^{2}}\frac{A}{4},
\end{equation}
where $L_{f}=\sqrt{\frac{e\lambda}{2}}$ and $\zeta\left(  3\right)  =%
{\displaystyle\sum\limits_{n=1}^{\infty}}
\frac{1}{n^{3}}\approx1.202$. In both cases, the area law in the brick wall
model is preserved after the GUP corrections to $n\left(  \omega,l\right)  $
are considered. Our result shows that the area law is still valid after the
GUP corrections to both $n\left(  \omega,l\right)  $ and the Hawking
temperature $T$ are included. If we choose the invariant cutoff $\varepsilon
=\frac{m_{p}}{\sqrt{90\pi}}$ in t' Hooft's original calculation, then the
dominant entropy term becomes the Bekenstein-Hawking entropy, $S\sim
S_{BH}\equiv\frac{A}{4m_{p}^{2}}$. Similarly, if we have $L_{f}=\sqrt
{\frac{17}{450\pi}}m_{p}\approx0.1m_{p}$ in our calculation, $S_{BH}$ also
appears for the leading entropy term. On the other hand, it is normally
assumed that the minimal $L_{f}$ is of the order of $m_{p}$. If $L_{f}\sim
m_{p}$, one finds the entanglement entropy for one scalar field $S\sim
10^{-2}S_{BH}$. Since the entanglement entropy depends on the number and
nature of quantum fields while $S_{BH}$ is universal, the \textquotedblleft
species problem\textquotedblright\ arises when the the entanglement entropy
makes important contributions to $S_{BH}$. Taking $L_{f}\sim m_{p}$ would
provide a resolution to the species problem.

The subleading logarithmic part of the entanglement entropy is also calculated
in our paper. It turns out that the subleading logarithmic part is universally
given by $-\frac{\kappa r_{h}}{45}\left(  1-\kappa\eta r_{h}\right)  \ln
\kappa\varepsilon$ in scenarios both with and without the GUP. For the
Schwarzschild metric with $f\left(  r\right)  =1-\frac{2M}{r}$, the subleading
logarithmic part becomes $-\frac{\ln\kappa\varepsilon}{45}$, which was first
obtained in \cite{CON-Solodukhin:1994yz}. On the other hand, we can estimate
the entropy of the black hole using eqn. $\left(  \ref{HKT1}\right)  $. \ In
fact, the angular momentum of the particle $L\sim pr_{h}\sim\omega r_{h}$ near
the horizon of the the black hole. Thus, one can rewrite $T$%
\begin{equation}
T\sim\frac{T_{0}}{1+\frac{2\omega^{2}}{M_{f}^{2}}}, \label{eq:Temp}%
\end{equation}
where $T_{0}=\frac{\hbar}{8\pi M}$ for the Schwarzschild black hole. As
reported in \cite{CON-AmelinoCamelia:2005ik}, the authors obtained the
relation $\omega\gtrsim\frac{\hbar}{\delta x}$ between the energy of a
particle and its position uncertainty in the framework of GUP. Near the
horizon of the the Schwarzschild black hole, the position uncertainty of a
particle will be of the order of the Schwarzschild radius of the black hole
\cite{CON-Bekenstein:1973ur} $\delta x\sim r_{h}$. Thus, one finds for $T$%
\begin{equation}
T\sim\frac{T_{0}}{1+\frac{m_{p}^{4}}{2M^{2}M_{f}^{2}}}, \label{eq: Temp}%
\end{equation}
where we use $\hbar=m_{p}^{2}$. Using the first law of the black hole
thermodynamics, we find the corrected black hole entropy is%
\begin{equation}
S=\int\frac{dM}{T}\sim\frac{A}{4m_{p}^{2}}+\frac{4\pi L_{f}^{2}}{m_{p}^{2}}%
\ln\left(  \frac{A}{16\pi}\right)  , \label{eq:entropy}%
\end{equation}
where $A=4\pi r_{h}^{2}=16\pi M^{2}$ is the area of the horizon. The
logarithmic term in eqn. $\left(  \ref{eq:entropy}\right)  $ is the well known
correction from quantum gravity to the classical Bekenstein-Hawking entropy,
which have appeared in different studies of GUP modified thermodynamics of
black holes\cite{CON-Bina:2010ir,CON-Chen:2002tu,CON-Xiang:2009yq}. Comparing
eqn. $\left(  \ref{eq:EntropyD}\right)  $\ with eqn. $\left(  \ref{eq:entropy}%
\right)  $, we note that there are two discrepancies in the subleading
logarithmic term, one of which is the sign and the other dependence on the
minimal length $L_{f}$. These discrepancies would imply that the entanglement
entropy could not solely account for the entropy of the black hole.

It has been shown in
\cite{IN-Li:2002xb,IN-Kim:2006rx,IN-Kim:2007bx,IN-Kim:2007nh,IN-Kim:2007if}
that the artificial cutoff parameter in the brick wall model located just
outside the horizon can be avoided if GUP is considered. However, this is not
the case in our paper. Actually, if one attempts to let $r_{\varepsilon}%
=0\,$\ in eqn. $\left(  \ref{eq:EntropyDivergentR}\right)  $, the entropy
diverges and a brick wall is still needed. How can we reconcile the
contradiction? One might note that we calculate the entanglement entropy to
$\mathcal{O}\left(  \frac{1}{M_{f}^{2}}\right)  $. For the typical energy
$\omega\sim T_{0}=\frac{\hbar\kappa}{2\pi}\,$, one finds that the
$\mathcal{O}\left(  \frac{1}{M_{f}^{2}}\right)  $ GUP corrections to the
entropy $\sim\frac{\omega^{2}}{f\left(  r\right)  M_{f}^{2}}\sim\frac
{\hbar^{2}\kappa^{2}}{4\pi^{2}f\left(  r\right)  M_{f}^{2}}$. At the wall at
$r_{\varepsilon}=r_{h}+2\kappa L_{f}^{2},$ we have $\frac{\hbar^{2}\kappa^{2}%
}{f\left(  r_{\varepsilon}\right)  M_{f}^{2}}\sim\frac{\hbar^{2}\kappa^{2}%
}{16\pi^{2}\kappa^{2}L_{f}^{2}M_{f}^{2}}\sim\frac{1}{16\pi^{2}}$. Thus, our
perturbative method is valid outside the wall at $r_{\varepsilon}%
=r_{h}+2\kappa L_{f}^{2}$. However, the perturbation would break down deep
within the wall and one needs to consider higher order contributions. The
divergence of the entropy eqn. $\left(  \ref{eq:EntropyDivergentR}\right)  $
as $r_{\varepsilon}\rightarrow0$ is more like due to the breaking down of our
perturbative method. In
\cite{IN-Li:2002xb,IN-Kim:2006rx,IN-Kim:2007bx,IN-Kim:2007nh}, it is crucial
for $\lambda\omega^{2}/f$ to be in the denominator of the integrand in eqn.
$\left(  \ref{eq:NLamda}\right)  $ to get rid of the wall. If one replaces
$\left(  1+\lambda\omega^{2}/f\right)  ^{-3}$ with $1-3\lambda\omega^{2}/f$ in
eqn. $\left(  \ref{eq:NLamda}\right)  ,$ the integral diverges as
$r\rightarrow r_{h}.$ The divergence arises simply because the Taylor
expansion of $\left(  1+\lambda\omega^{2}/f\right)  ^{-3}$ breaks down when
$\lambda\omega^{2}/f>1$.

In this paper, we calculate the entanglement entropy of a massless scalar
field near the horizon of a 4D spherically symmetric black hole in the brick
wall model incorporating the minimal length effects. We show that the leading
term of the entropy is proportional to the horizon area. If the minimal length
$L_{f}\sim m_{p}$, the entanglement entropy makes a small contribution to the
Bekenstein-Hawking entropy, which might resolve the \textquotedblleft species
problem\textquotedblright. The subleading logarithmic term is also calculated.
The result is the same as the one in the usual brick wall model without the
minimal length and independent of the minimal length $L_{f}$.\ 

\noindent\textbf{Acknowledgements. }

We would like to acknowledge useful discussions with Y. He, Z. Sun and H. W.
Wu. This work is supported in part by NSFC (Grant No. 11005016, 11175039 and
11375121) and SYSTF (Grant No. 2012JQ0039).

\end{document}